\documentclass[runningheads]{llncs}
\usepackage{soul}
\usepackage{graphicx}
\usepackage[disable]{todonotes}

\usepackage{comment}
\usepackage{placeins} 
\usepackage[]{mdframed}
\newcommand{\framedtext}[1]{%
\par%
\noindent\fbox{%
    \parbox{\dimexpr\linewidth-2\fboxsep-2\fboxrule}{#1}%
}%
}
\begin{document}
\title{Norm Violation Detection in Multi-Agent Systems using Large Language Models}
\subtitle{A Pilot Study}
\titlerunning{Norm Violation Detection using LLMs}

\author{Shawn He\inst{1} \and
Surangika Ranathunga\inst{1}\orcidID{0000-0003-0701-0204} \and \\
Stephen Cranefield\inst{2}\orcidID{0000-0001-5638-1648} \and \\
Bastin Tony Roy Savarimuthu\inst{2}\orcidID{0000-0003-3213-6319}
}

\authorrunning{He et al.}
%
\institute{Massey University, Auckland, New Zealand\\
\email{Shawn.He.1@uni.massey.ac.nz, S.Ranathunga@massey.ac.nz}\\
 \and
University of Otago, Dunedin, New Zealand\\
\email{\{stephen.cranefield,tony.savarimuthu\}@otago.ac.nz}
}

\maketitle              

\begin{abstract}
  Norms are an important component of the social fabric of society by prescribing expected behaviour. In Multi-Agent Systems (MAS), agents interacting within a society are equipped to possess social capabilities such as reasoning about norms and trust. Norms have long been of interest within the Normative Multi-Agent Systems community with researchers studying topics such as norm emergence, norm violation detection and sanctioning. However, these studies have some limitations: they are often limited to simple domains, norms have been represented using a variety of representations with no standard approach emerging, and the symbolic reasoning mechanisms generally used may suffer from a lack of extensibility and robustness. In contrast, Large Language Models (LLMs) offer opportunities to discover and reason about norms across a large range of social situations. This paper evaluates the capability of LLMs to detecting norm violations. Based on simulated data from 80 stories in a household context, with varying complexities, we investigated whether 10 norms are violated. For our evaluations we first obtained the ground truth from three human evaluators for each story. Then, the majority result was compared against the results from three well-known LLM models (Llama 2 7B, Mixtral 7B and ChatGPT-4). Our results show the promise of ChatGPT-4 for detecting norm violations, with Mixtral some distance behind. Also, we identify areas where these models perform poorly and discuss implications for future work.         

\end{abstract}

\section{Introduction}

Norms prescribe behaviours that are expected by agents in a society. Norms are pivotal to establishing social order within a society as they facilitate coordination and cooperation between participating agents \cite{bicchieri2014}. A software agent that is \emph{norm capable} should be able to identify existing norms, detect situations where norms are violated and plan to either comply with norms or choose to violate them with an understanding of the likely consequences. These capabilities (and more) are the subject of the research field of normative multi-agent systems.

This paper focuses on norm violation detection, also known as norm monitoring. The importance of this ability in human society is evidenced by our innate ability to detect norm violations. Brain studies in humans using EEG recordings show a negative deflection of electrocortical potential that occurs at around 400ms, which shows they have recognised the violation of expectations, including those that arise from norms \cite{salvador2020}. Without a robust mechanism to detect norm violations, effective norm enforcement (e.g., by sanctioning) is not possible, undermining the spreading and emergence of norms and reducing coordination between agents.

Prior normative MAS research has a number of limitations. Norms are encoded and formalised in a variety of representations with no standard approach emerging. We believe this slows progress in the field as advances made by one researcher are not easily exploited by others. Also, norm representations and reasoning mechanisms are often developed for and evaluated in simple static problem domains. Furthermore they are predominantly based on the symbolic AI paradigm, and may therefore lack robustness when deployed in domains outside their initial design constraints~\cite{Minsky_1991}. Thus, current techniques may have limited applicability to the types of domain-independent sociotechnical system that are enabled today by smartphones, social media networks and personal assistants (which users are rapidly becoming accustomed to since the advent of ChatGPT).

Large Language Models (LLMs) offer promise in addressing these issues since LLMs can work with norms expressed in natural language rather than committing to one of the many norm representations in the literature. Recently, Natural Language Processing (NLP) researchers have started to investigate LLM capabilities for normative reasoning. For example, pre-trained language models already have been shown to possess some notion of norms, and they can be further trained with publicly available norm-related datasets (discussed in Section~\ref{sec:norm_datasets} below). However, in that research, normative reasoning has been done on text such as chatbot conversations, narratives or situation descriptions. In contrast, in an MAS, an agent must deliberate on norm violation based on a sequence of events it perceives from its environment. How an LLM would be capable of determining norm violations from such an event sequence has not been investigated before. This work aims to bridge this gap.

In this work, we evaluated the norm identification capability of commonly used public and proprietary LLMs. We prompted them with a sequence of events that occurred in a simulated environment related to a household setup, along with a set of norms defined in that environment. The LLMs were asked to determine whether each norm was violated or not, in a given scenario. Comparing these results with human-generated answers we found that the capabilities of different LLMs for norm violation detection varied, but the best model, ChatGPT-4, identified norm violations with an accuracy of 86\%. However, the performance depended on the type of norm. 

The rest of the paper is organised as follows. Section 2 discusses prior work related to norm violation detection in MAS and NLP research. Section 3 discusses our experimental setup and Section 4 provides the results. We provide a detailed discussion of our results in Section 5. Finally, Section 6 concludes the paper.

\section{Prior work}

\subsection{MAS Research on Norm Violation Detection}
In the MAS literature, norms are expressed using a variety of notations such as conditional rules, temporal logic constraints and action description languages (such as the event calculus). A short survey is given by Cranefield et al.~\cite[Sec.~2]{Cranefield_interdependent_exps_2011}. Various abstract models of norms have also been proposed that specify separate components of a norm such as the type (obligation, etc.), triggering condition, the normative constraint, expiration condition and sanction.
There are also a host of ad hoc domain-specific notations for norms.

Given this diversity, it is not surprising that many mechanisms for norm violation detection have been proposed. These include those that arise naturally from the underlying formalisms such as forward-chaining rule engines, temporal projection in the event calculus and temporal logic model checking as well as the design of separate entities to monitor norms (see Dastani et al.\@ for a survey of the latter approach \cite{dastani_norm_monitoring_2018}).

\subsection{NLP Research for Norm Violation Detection}
In NLP, instead of norm violation detection, researchers focus on predicting `moral judgement' in a given situation~\cite{jiang2021can,ziems2023normbank}. The common approach to training such a prediction system is to fine-tune a pre-trained language model with $situation\_description$, $moral\_judgement$ pairs. Here, the $moral\_judgement$ refers to whether an action was normative or not. This kind of model has been commonly implemented as a text classification system, where a textual description of an action gets classified as normative or not~\cite{nahian2020learning}.

\label{sec:norm_datasets}
More recent work has produced models that are capable of more fine-grained reasoning. For example, \textsc{NormBank}~\cite{ziems2023normbank} categorises a given situation as \textit{ok, expected} or \textit{unexpected}. The publicly available Delphi system\footnote{\url{https://delphi.allenai.org/}}
\cite{jiang2021can} responds with one of the following judgements: \textit{rude, ok, bad, wrong, expected, shouldn't}, when prompted with a situation (expressed in a single sentence). 
While Delphi carries out normative reasoning in isolation with no reference to surrounding context, ClarifyDelphi~\cite{pyatkin2023clarifydelphi} is capable of asking questions to elicit salient context.
While all this research made use of pre-trained language models, none has investigated the capabilities of the recently released LLMs, which we are using in this research. Moreover, this NLP research focused on norm violation present in textual descriptions such as narratives or chatbot output. In contrast, we are focusing on an event sequence received by an agent.

\subsection{LLM Agents}

The field of LLM agents is a rapidly advancing thread of research that considers how intelligent software agents can leverage the abilities of LLMs to help them make decisions. Recent survey papers by Wang et al.~\cite{wang2023survey} and Xi et al.~\cite{xi2023rise} give an overview of this field. A key benefit is that an LLM can provide an agent with the ability to operate in a wide range of environments unlike traditional approaches to agent development that are usually restricted to specific problem domains.

There has been prior research on how LLM-based agents can interact in a society~\cite{xi2023rise}, including mechanisms that enable agents to exhibit ``believable individual and emergent social behaviors'' \cite{Park2023GenerativeAgents}. However, prior work has not considered how LLMs could assist software agents with social reasoning, including norm awareness and monitoring.

A companion paper \cite{BlueSkyPaperOURArchive} discusses the above points in more detail and proposes a number of potential paths to extending LLM agents to become \emph{normative} LLM agents.

\section{Methodology}
\subsection{Simulating an Agent Environment}

We used the grammar of Nematzadeh et al.~\cite{nematzadeh2018evaluating} to create stories that describe a sequence of actions performed by a set of agents in a pre-defined environment. It has been used to test Theory of Mind (ToM) capabilities of LLMs \cite{nematzadeh2018evaluating,alechina2018}. This grammar is defined in terms of a set of \textit{entities} (i.e., objects that may have \textit{properties}, and agents) and a set of \textit{predicates}. A predicate takes an entity as its subject or object. A predicate can be an action that an agent carries out (e.g. move, enter). Each story is generated in such a manner that it constitutes of a set of tasks. In each task, several agents (usually two) carry out different actions such as entering or exiting rooms or moving things around. In addition to the agent actions, the stories may contain random events that represent noise.

We introduced the following modifications to the implementation of Nematzadeh et al.~\cite{nematzadeh2018evaluating} to generate more complex scenarios:
\begin{enumerate}
    \item In their work, tasks in a story are arranged one after the other, meaning that the agent would perceive actions of one agent before perceiving the other's.  However, this simple scenario is not what would happen in an agent environment---an agent can perceive actions of different agents in an interleaved fashion. In order to achieve this realistic scenario, we interchanged actions of different agents using a topological sort algorithm~\cite{kahn1962}.
    \item Their stories have only one random event (\textit{phone rang}). We added two more events  (\textit{kettle whistled}, and \textit{cat meowed}), and randomly added them as noise to the generated stories. 
    \item We generate stories by varying the number of tasks from one to four. This results in stories that have the number of events ranging from 6 to 30.
\end{enumerate}

\begin{figure} [h]
\centering
    \includegraphics[width=0.8\linewidth]{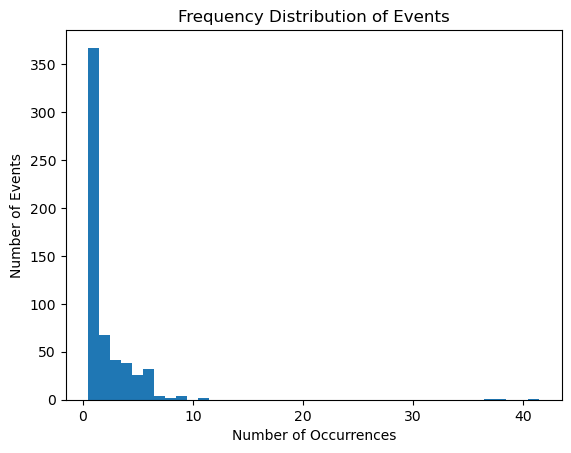}
    \caption{Frequency distribution of events across the 80 stories}
    \label{fig:freq_graph}
\end{figure}

We used the same set of agents and objects that were defined in the original implementation. Altogether, we generated 80 stories, where every 20 stories had a different number of tasks (i.e., 20 stories each for one, two, three and four tasks). Figure~\ref{fig:story} shows a sample story (with four tasks) generated from the modified version of the implementation of Nematzadeh et al.~\cite{nematzadeh2018evaluating}. Figure~\ref{fig:freq_graph} shows the number of occurrences of each event across the 80 stories. This figure confirms that most stories are unique. To be specific, out of a total of 1317 events across the 80 stories, 367 are unique events and only 219 are duplicated events.

\begin{figure}[t]
\framedtext{
Alexander entered the sunroom.\\
Alexander moved the clock to the black suitcase.\\
Peter entered the sunroom.\\
Peter exited the sunroom.\\
Peter entered the study.\\
Peter exited the study.\\
Emily entered the sunroom.\\
Emily moved the fork to the blue bucket.\\
Emily exited the sunroom.\\
Phone rang.\\
Alexander exited the sunroom.\\
Emily entered the study.\\
Emily moved the tomato to the white crate.\\
Emily exited the study.\\
Peter entered the basement.\\
The hat is in the black bottle.\\
The fork is in the black suitcase.\\
Phone rang.\\
The clock is in the blue bucket.\\
Peter moved the hat to the red carpet.\\
Peter exited the basement.\\
Emily entered the sunroom.\\
Emily exited the sunroom.\\
Emily entered the basement.\\
The tomato is in the green refrigerator.\\
Emily exited the basement.
}
\caption{An example story used in our experiment}
\label{fig:story}
\end{figure}

\begin{figure}[t]
\framedtext{
\small
\emph{Generic norms:}
\begin{enumerate}
    \item [1] - You should not enter an occupied bathroom. (PN)
    \item [2] - Vegetables and fruits (except bananas) should be kept in the refrigerator. (ON)
    \item [3] - Before using the staircase, you should wait until it is free. (ON)
    \item [4] - If sharp objects are present in the household, they should be kept out of reach of children. (ON)
    \item [5] - If the phone rings, the person closest to the phone must pick it up. (ON)
    \item [6] - There should be no fruits and vegetables containing citric acid inside the house. (PN)
\end{enumerate}

\emph{Role-based norms:}
\begin{enumerate}
    \item [7] - Kids should not enter crawl spaces. (PN)
    \item [8] - You should not enter a couple’s bedroom without knocking. (PN)
    \item [9] - Kids should be supervised inside the workshop all the time. (ON)
    \item [10] - Father should not be disturbed while he is in his study room. (PN)
\end{enumerate}
}
\caption{The norms used in our experiment}
\label{fig:norms}
\end{figure}

\subsection{Defining Norms}
The list of objects and agents in the simulation suggest that Nematzadeh et al.~\cite{nematzadeh2018evaluating} have considered events taking place in a home. Therefore we defined 10 norms that can be realistically applied in a household setup. We defined a home with four agents: mother (Emily) and father (Peter) with two young children (Alexander, who is a 5-year-old boy and Ann, who is a 7-year-old girl). The set of norms we used is shown in Figure~\ref{fig:norms}. We defined two types of norms: generic norms (6 norms: 1 to 6) and role-based norms (4 norms: 7 to 10). They are further divided into Obligation Norms (ON) and Prohibition Norms (PN) as indicated at the end of each norm in Figure~\ref{fig:norms}.


\begin{figure}[!t]
\framedtext{
\fontsize{7.5}{8.5}\selectfont

There is a family that adheres to certain norms, listed below from Norm 1 to 10. The family consists of Emily (wife), Peter (husband), and their children, Alexander (5-year-old boy) and Ann (7-year-old girl).\\

As a sophisticated assistant, you are tasked with analyzing a story related to this family to identify any violations of the given norms. Your goal is to systematically break down the analysis for each norm, ensuring a thorough evaluation based on the events listed in the story.\\

The family's story unfolds with the whole family initially being in the living room, alongside a phone.\\

There is no other person in the house.\\

At the beginning of the story, no norms are violated.\\

If an event is not mentioned in the story, then it did not occur.\\

Your job is to evaluate each event in the story against the below given norms to identify any violations.\\

Norms to Evaluate:\\
$\{$norms$\}$\\
 
Below is the story about the family's events.\\

Story:\\

$\{$story$\}$\\ 

Your evaluation should consider each norm individually. For each norm, provide a clear response in the following format:\\

- Norm 1\\

- Violation: [Choose one of three options given below]\\

- No: The norm was not violated or its irrelevance was given the story.\\

- Cannot be determined: There isn't clear evidence to determine whether the norm was violated or not.\\

- Yes: The norm was violated.\\

- Reasoning: [Explain your rationale here, referencing specific events in the story as evidence.]\\ 

Please proceed with your analysis, ensuring a thoughtful and comprehensive examination of each norm as it relates to the story's events.\\
}
\caption{Our prompt template for ChatGPT-4}
\label{fig:prompt}
\end{figure}

\subsection{Prompting LLMs and Human Evaluation}
We used ChatGPT-4 as our proprietary LLM and Llama 2 (Llama-2-7b-chat-hf) and Mixtral (Mixtral-8x7B-Instruct-v0.1) as the open source LLMs. The two open source models were selected considering their promising performance despite the smaller model size. After a preliminary set of experiments and discussions had within the research team over several rounds, the prompt shown in Figure~\ref{fig:prompt} was used in the final experiments. The figure shows the ChatGPT-4 version. Prompts for other two models are more or less the same, except for the model-specific tags. Each of the 80 stories we generated were provided as prompts to each of the LLMs along with the ten norms investigated for each story. 

We also carried out a human evaluation, in order to obtain the ground truth. The same 80 stories were provided to external human evaluators, along with the scenario description and the set of norms. For each story and each of the norms, they were asked to decide among the following three possibilities: Norm violated, norm not violated, or not enough information to give a judgement. Each story was evaluated by three human annotators. Altogether we used nine annotators for the evaluation. Six were Computer Science undergraduates and three were recent Computer Science graduates. Two were females and the rest were males. Inter-annotator agreement for the human evaluation was conducted. The Fleiss Kappa value was 0.6826, which indicates a substantial agreement.  

\section{Results}
From the human annotated data, we used majority voting to select the decision corresponding to a norm with respect to a given story. We evaluated model performance using human annotation as the ground truth. However, despite being given the option to decide between three options, yes/no/cannot be determined, humans have always selected one of the two options: yes or no. Therefore, if a model has the output `cannot be determined', we map that output to `no'. 

Table~\ref{tab:overall_result} shows the performance (accuracy) of each LLM against human judgement. The result indicates that ChatGPT-4 has the overall best performance, with 86\%. Mixtral, the second best technique, falls behind by a slight margin. The result also indicates that the model performance is not consistent across norms - i.e., for some norms it performs much better (e.g., for Mixtral, 97.5\% for norm 9 when compared to 28.75\% for norm 2). Therefore we conducted further analysis based on norm category (whether the norm is generic or role-based) and also based on the type of the norm considered (whether the norm is a prohibition norm or an obligation norm). Result in Table~\ref{tab:norm_category_result} shows that all the models struggle to detect generic norms than role-based norms. Result in Table~\ref{tab:norm_type} shows that all the three models were able to detect prohibition norms better than obligation norms (82\% vs. 68\%).

\begin{figure} [h]
\centering
    \includegraphics[width=0.8\linewidth]{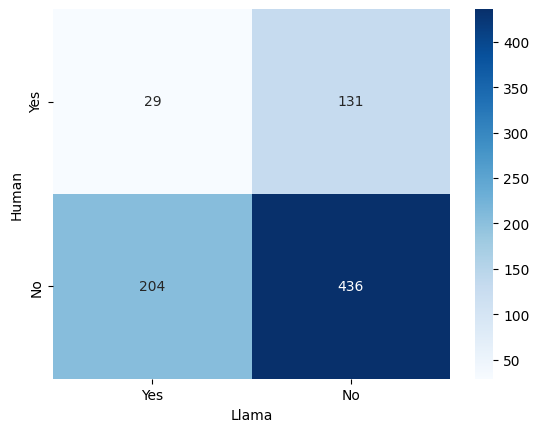}
    \caption{Confusion Matrix for Llama 2 result against ground truth}
    \label{fig:confusion_matrix}
\end{figure}

\begin{figure} [h]
\centering
    \includegraphics[width=0.8\linewidth]{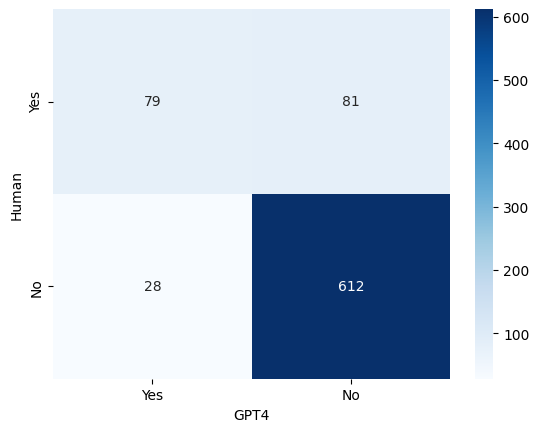}
    \caption{Confusion Matrix for ChatGPT-4 result against ground truth}
    \label{fig:confusion_matrix_gpt}
\end{figure}

We also calculated the disagreement between ground truth and each of the models in determining norm violation. The result for the worst-performing model (Llama 2) is shown in Figure~\ref{fig:confusion_matrix} in the form of a confusion matrix. This model seems to struggle in identifying norm violations, more than determining that there is no violation. In other words, out of the instances where a human identified a norm violation, 81\% of the time the model has not been able to identify this violation. On the contrary, when the human determines a non-violation, the model's decision differs only by 31.8\%. Figure~\ref{fig:confusion_matrix_gpt} shows the confusion matrix for the best performing model ChatGPT-4. For this model, out of the instances where a human identified a norm violation, only 51\% of the time the model has not been able to identify this violation. When the human determines a non-violation, the model decision differs only by 4\%. This further highlights the superior results of ChatGPT-4.
\begin{table}[t]
\caption{Overall accuracy (in percentages) for the three models for the 10 norms (N1 to N10)}
\label{tab:overall_result}
\centering
\fontsize{8}{9}\selectfont
\begin{tabular}{| l | r | r | r | r | r | r | r | r | r | r | r|}
\hline
\bf Model & \bf N1 & \bf N2 & \bf N3 & \bf N4 & \bf N5 & \bf N6 & \bf N7 & \bf N8 & \bf N9 & \bf N10 & \bf Average\\
\hline
Llama 2 & 43.75 & 21.25 & 92.50 & 37.50 & \textbf{70.00} & 58.75 & 95.00 & 17.50 & 93.75 & 51.25  & 58.0 \\
\hline
Mixtral & 97.50 & 28.75 & 90.00 & 85.00 & 68.75 & \textbf{93.75} & \textbf{96.25} &\textbf{ 96.25 }& \textbf{97.50} & 51.25 &81.0 \\
\hline
ChatGPT-4 & \textbf{100.00} & \textbf{62.50} & \textbf{96.25} & \textbf{87.50 }& \textbf{70.00} & 92.50 & 93.75 & \textbf{96.25} & 93.75 &\textbf{ 71.25 } & \textbf{86.0}\\
\hline

\end{tabular}
\end{table}

\begin{table}[t]
\caption{Accuracy results (in percentages) based on norm category}
\label{tab:norm_category_result}
\centering

\begin{tabular}{|l| l| l| l|l|}
\hline
\textbf{Norm category} & \textbf{ChatGPT-4} & \textbf{Mixtral} & \textbf{Llama 2} & \textbf{Average}\\ 
 \hline
Role-based norm & 94.06 & 95.94 & 66.25 & 85.42\\
\hline
Generic norm & 81.25 & 70.21 & 52.71 & 68.05\\
\hline
\end{tabular}
\end{table}

\begin{table}[!t]
\caption{Accuracy results (in percentages) based on norm type}
\label{tab:norm_type}
\centering
\begin{tabular}{|l| l| l| l| l|}
\hline
 \textbf{Norm type} & \textbf{ChatGPT-4} & \textbf{Mixtral} & \textbf{Llama 2} & \textbf{Average}\\
 \hline
Prohibition Norm (PN) & 90.25 & 87.25 & 68.50 & 82.00\\
\hline
Obligation Norm (ON) & 82.50 & 73.75 & 47.75 & 68.00\\
\hline

\end{tabular}
\end{table}

\section{Discussion}

We make several observations based on the results. First, ChatGPT-4 performed better than the other models with an accuracy of 86\%. This result is not surprising given that GPT Turbo, the model used in ChatGPT-4 topped the leaderboard for general reasoning at the time of writing~\cite{fan2023}. However, we note that the result obtained is far from being perfect. Software agents performing actions based on the result even from this best performing LLM might still be performing norm-violating actions based on the advice received, and hence may be subject to sanctions. However, note that in this work we simply prompted the LLMs in a zero-shot manner (i.e.~without showing any examples to the LLM, either via in-context learning or fine-tuning). Therefore, we suggest the creation of fine-tuned or RAG (Retrieval Augmented Generation)-based models to further improve norm detection ability of LLMs. Use of advanced reasoning techniques such as Tree-of-Thought prompting~\cite{yao2024tree} may further improve outcomes. More broadly, norm competence of LLMs \cite{BlueSkyPaperOURArchive}, of which norm violation detection is only one aspect, should be improved.

Second, we observed that for a few norms all the LLMs performed poorly (Llama 2 7B and Mixtral 7B, more so than ChatGPT-4). For example, violations of norm 2 about keeping vegetables and fruits in the refrigerator have been identified poorly. Perhaps, it could be that this norm may have been in conflict with common knowledge reasoning where only green vegetables are kept in the refrigerator. In some cases, the relatively poor result of Llama 2 may have been because of the non-availability of data. For example, norm 6 requires the ability to determine whether a vegetable or fruit contains citric acid (e.g., beetroot in one story). However, it may be difficult for the LLM to identify such information. There are limited sources on the Internet that may have this information so it may not be in the LLM's training data. Also, as we did not provide a floor plan of the house to the LLM, its performance may have been impacted. For example, we noticed that in certain instances, Llama 2 made incorrect inferences related to various rooms of the house (e.g.~the kitchen) having crawlspaces. 

Third, we found that prohibition norm violations were more successfully identified by LLMs than obligation norm violations. This may be because there may be more instances of prohibition norms than obligation norms in the dataset to pre-train the LLM, since prohibitions are more commonly stated than obligations.

Fourth, our work shows that certain LLMs were better at identifying non-violation of norms in a story than identifying violations (e.g., for Llama 2 this difference is about 50\%). This may be because the dataset in which these LLMs are pre-trained are not balanced (i.e., imbalance between violations and non-violations). This suggests these models need may to be trained with more data with violations.

Fifth, while most of the human-coded data contained only yes and no classifications, indicating violations and non-violations respectively, two of the models (Mixtral and ChatGPT-4) contained results that were more nuanced. In many cases ChatGPT-4 returned `cannot be determined'---an option that was available to the humans also, which implies there is not enough evidence to determine whether a norm was violated. Also Mixtral  returned `not applicable' (which could mean a conditional norm was not triggered and therefore not violated). For example, for the norm that prohibits the second person entering the staircase when someone is already on it, if the story does not involve a staircase, it returned the result `not applicable' since the staircase was not mentioned. This was a surprising result since we had provided only \emph{yes}, \emph{no} and \emph{cannot be determined} as the three options for the language model. In future work, we can consider  providing four answer options: yes, no, not applicable and cannot be determined.

Sixth, we observed that the norm identification capability of an LLM depends on the prompt. In fact, prompt brittleness has been highlighted as a limitation of LLMs~\cite{zamfirescu2023johnny}. As mentioned earlier, we had to experiment with multiple prompts before reaching the most promising prompt. For example, during our initial experiments, we identified that the LLM is not capable of determining the gender of a character from their name. Therefore, we explicitly had to mention who is the husband, wife and children etc. are. Also, the LLMs seemed to get confused with the location of characters. Therefore we had to explicitly mention that all the characters are in the living room at the beginning of the story. Some LLMs tended to assume that the story may contain other characters that have not been mentioned in the story. Therefore we explicitly had to mention that there are no other characters in the house. In order to reduce LLM hallucination, we had to explicitly state that if an event is not mentioned in a story, that event did not occur. Additionally, explicitly stating the expected format of the LLM output was also needed.

Seventh, while this work considered both generic and role-specific norms, these are relatively simple norms. In the future, we intend to experiment with more complex norms where there are several conditions involving multiple entities.

Eighth, this work does not propose how norm violation detection information is used by agents for decision-making (e.g., to punish other agents), which can form the focus of future work. 

Finally, this work is a pilot study on investigating normative reasoning focusing on one aspect, norm violation detection. We encourage other researchers to pursue other aspects of normative reasoning as suggested in our companion discussion paper~\cite{BlueSkyPaperOURArchive}. Additionally, this paper provides a comparison of norm violation detection ability of LLMs against human evaluations. In the future, we will investigate how well LLMs perform when compared to logic-based approaches. Having said that there has been prior work that have considered neuro-symbolic approaches where textual descriptions of norms are converted to logic~\cite{ferraro2020,chaudhury2021,zhan2023} and vice-versa~\cite{calo2022} which can be employed here. Also, LLMs themselves can be used for this purpose (i.e., convert text to logic \cite{borazjanizadeh2024}). For our current experiments, we used a simple simulation. However, there have already been studies to deploy autonomous agents in more complex simulations~\cite{ranathunga2011interfacing}. Therefore in the future, we intend to experiment with more complex simulations. 

\section{Conclusion}
This work aimed at investigating the effectiveness of LLMs in detecting norm violations based on 80 stories. Of the three LLMs considered in the work (Llama 2 7B, Mixtral 7B and ChatGPT-4), we observed that ChatGPT-4 offered the most promise, with 86\% accuracy. We also observed the LLMs detected violations of prohibition norms with higher accuracy than those of  obligation norms. Additionally, violations of role-based norms were more accurately identified than those of generic norms. Also, LLMs were better at identifying when norms weren't violated than when they were violated. These results demonstrate that, while LLMs show promise in norm violation detection, there is scope for creating better models for detecting norm violations. We believe such enhanced models will be beneficial for recommending actions to be taken by an agent within a multi-agent system, īn order to create norm-conforming agent societies where potential norm violations are effectively identified and punished. In this study, we focused only on norm violation detection, which is an example of the normative reasoning capability of an agent. However, we believe that LLMs can be leveraged to achieve other norm competencies in agents, such as norm discovery and norm conformance.

\section{Acknowledgements}
We would like to thank the following individuals who volunteered  for the human evaluation: Dilan Nayanajith, Paul Janson, Sanjeepan Sivapiran, Kumuthu Athukorala, Yasith Heshan, Rumal Costa, Thevin Senath, Harshani Bandara and Nimesh Ariyarathne.

This research was funded by a research grant from the College of Sciences, Massey University, New Zealand.

\bibliographystyle{splncs04}
\bibliography{coine2024}
\end{document}